\documentstyle[12pt,epsf]{article} 
\newcommand{\VV}{{\cal V}}

\newcommand{\wt}{\widetilde}

\newcommand{\be}{\begin{equation}}
\newcommand{\ee}{\end{equation}}
\newcommand{\ben}{\begin{eqnarray}\displaystyle}
\newcommand{\een}{\end{eqnarray}}
\newcommand{\refb}[1]{(\ref{#1})}
\newcommand{\p}{\partial}

\begin{document}

{}~
\hfill\vbox{\hbox{hep-th/0009090}\hbox{MRI-P-000903}
}\break

\vskip 2.0cm

\centerline{\large \bf Uniqueness of Tachyonic Solitons}

\vspace*{6.0ex}

\centerline{\large \rm Ashoke Sen}

\vspace*{6.5ex}

\centerline{\large \it Mehta Research Institute of Mathematics}

\centerline{\large\it
and Mathematical Physics, Chhatnag Road,}

\centerline{\large \it   Jhoosi,
Allahabad 211019, INDIA}
\vspace*{1ex}
\centerline{E-mail: asen@thwgs.cern.ch, sen@mri.ernet.in}

\vspace*{4.5ex}

\centerline {\bf Abstract}
\bigskip

It has been conjectured that condensation of tachyons on a bosonic D-brane
gives rise to vacuum / soliton solutions which are independent of the
initial magnetic field on the D-brane. We present evidence for this
conjecture using results from two dimensional conformal field theory. In
particular we identify a continuous path in the configuration space of
open string fields which interpolates between D-brane configurations with
two different quantized magnetic flux. 

\vfill \eject

\baselineskip=17.3pt



The dynamics of tachyon, massless scalars and gauge fields on a
D-$p$-brane of 26 dimensional bosonic string theory in flat Minkowski
signature
space-time in static gauge is described by the
action
\be \label{e1} 
S_{BI} =-{1\over g_c} {1\over (2\pi)^p}
\int d^{p+1}x \, \VV(T)\sqrt{-\det\Big(g_{\mu\nu} + \p_\mu
Y^i\p_\nu Y^i + 2 \pi F_{\mu\nu}\Big)} + \ldots\, , 
\ee 
where $x^\mu$
denote the world-volume coordinates of the brane $(0\le\mu\le p)$, $y^i$
denote the space-time coordinates transverse to the brane $(p+1\le i\le
25)$, 
$ Y^i$
denote the massless scalar fields on the D-brane world-volume associated
with the coordinates $y^i$, 
$g_c$ is the closed string coupling constant, $g_{MN}$
is the closed string metric (which we take to be constant with
$g_{i\mu}=0$, $g_{ij}=\delta_{ij}$), 
$F_{\mu\nu}=\p_\mu A_\nu -\p_\nu A_\mu$ denote the gauge field strength,
$T$ denotes the tachyon field, and $\VV(T)$ denotes the tachyon potential
with the D-brane tension term included. $\VV(T)$ 
has a maximum at $T=T_{max}$ describing the original D-brane
configuration and has been conjectured to have a local minimum at
$T=T_{min}$ where
$\VV(T_{min})=0$\cite{9902105,SECOND}. 
We have chosen $\alpha'=1$ and have normalized
$\VV$ so that $\VV(T=T_{max})=1$. $\ldots$ denotes terms containing
derivatives
of $T$, $\p_\mu Y$ and $F_{\mu\nu}$. 

It has also been conjectured that the minimum of the potential $T=T_{min}$
describes the vacuum without a D-brane\cite{9902105}, and that at
this minimum all different values of $F_{\mu\nu}$ and $ Y^i$ actually
describe the same physical configuration in open string field
theory\cite{0009038}. It follows from this, using the techniques of
non-commutative field theory\cite{9711162,9903205,9908142}, that many
apparently
different soliton solutions, for which $T$ approaches $T_{min}$
asymptotically, but the asymptotic values of $F_{\mu\nu}$ are different,
also describe the same configuration\cite{0009038}. This in turn resolves
some of the
puzzles raised in refs.\cite{0007226,0008013} in interpreting
non-commutative tachyonic solitons as
D-branes\cite{0003160,0005006,0005031}. 

Some evidence for this conjecture comes from the string field theory
results of refs.\cite{0007153,0008033}.
In this paper we provide
an evidence for this conjecture using the techniques of two dimensional
conformal field theory (CFT).
For simplicity of notation, we shall take our initial system to be a D-2
brane, and assume that the directions tangential to the 2-brane have been
compactified on a torus $T^2$. In this case, the above conjecture has a
somewhat dramatic consequence. Since the total magnetic flux through $T^2$
is quantized, in terms of the field variables appearing in the
Born-Infeld action it is not possible to continuously deform a
configuration with a given magnetic flux to a configuration with a
different magnetic flux {\it even via off-shell field configurations}. On
the other hand if the conjecture is correct, then starting from a
configuration with a given magnetic flux at $T=T_{max}$, we can deform $T$
to $T_{min}$, and since at $T=T_{min}$ all magnetic field background
describes the same configuration, we can change the value of the magnetic
field to any other value allowed by the quantization rules. We can then
deform the tachyon field back to $T=T_{max}$. This gives a continuous path
in the configuration space of open string field theory interpolating
between D-branes with different amounts of magnetic flux. 

A T-dual version of this phenomenon is as follows. Let us take a
D1-brane, and take the direction $x^1$ tangential to the D-brane, and a
direction
(say $y^2$)
transverse to the D-brane to be compact. As usual, we denote by $ Y^2$
the massless scalar field on the D-brane world-volume associated with the
coordinate $y^2$.
Now consider a classical field configuration
on the D1-brane world-volume theory of the form:
\be \label{e2}
 Y^2 = a x^1\, ,
\ee
where $a$ is a constant.  In this case compactness of the 1 and 2
directions imply that only discrete values of $a$
are allowed. In particular, if both 1 and 2 directions have the same
periodicity (say $2\pi$), then $a$ must be an integer,\footnote{This
describes a D-string pointing along the vector $(1,a)$ in the $(x^1,y^2)$
plane for integer $a$. There are also other allowed D-string
configurations
pointing along the vector $(p,q)$ for any relatively prime pair of
integers $p,q$. They can be represented as classical configurations in
the world-volume theory of $p$ D-strings lying along $x^1$.} since as
$x^1$
changes by
$2\pi$, $ Y^2$ must change by an integer multiple of $2\pi$. Again, the
Born-Infeld action
describing the D-brane world-volume theory does not allow a continuous
deformation of fields which interpolates between
configurations with different values of $a$. However, if the conjecture
stated above is correct, then we should be able to interpolate between
these two
field configurations by starting with a given value of $a$, taking $T$ to
$T_{min}$ where all values of $a$ correspond to the same configuration,
and then changing $T$ back to $T_{max}$ with $a$ taking a different
value.

In this paper we explicitly demonstrate this phenomenon using the
techniques of two dimensional CFT. For definiteness we
shall focus on the problem of interpolating between two magnetic field
backgrounds in D2-brane wrapped on a torus. Since it is difficult to
describe tachyon condensation into vacuum using the CFT
techniques, one would have thought that even if the
interpolation outlined above had been possible, it will be difficult to
demonstrate this using CFT techniques since the path passes through the
vacuum configuration.  But as we shall see, we do not need to go all the
way down to the vacuum configuration for this interpolation; it is
possible to find a path via a codimension one soliton, {\it i.e.} a
D1-brane in this case. Since formation of a codimension one soliton via
tachyon condensation is a well understood process in conformal field
theory\cite{RECK,9902105,9406125,0003101}, we can use CFT techniques for
studying this
process. 

More specifically, we shall show that under certain conditions the initial
system of D2-brane with magnetic flux can be taken to a D1-brane via
marginal deformation, and furthermore, that this D1-brane can be taken to
a D2-brane without magnetic flux via a relevant deformation. In the
language of string field theory this implies that in the configuration
space of string fields there is a continuous path which interpolates
between a D2-brane configuration with magnetic flux and a D2-brane
configuration without magnetic flux. This is precisely what is expected
according to the conjecture stated earlier.

We
label the D2-brane world volume by coordinates $(x^0,x^1,x^2)$ with
$x^1\equiv x^1 + 2\pi$, $x^2\equiv x^2+2\pi$,
take the background anti-symmetric tensor field to be 0, and the
background closed string metric to be
\be \label{e3}
g_{\mu\nu} = \pmatrix{R_1^2 & \cr & R_2^2} \qquad \hbox{for} \quad
\mu=1,2\, ,
\ee
with the rest of the components of $g_{MN}$ being equal to those of the
Minkowski metric $\eta_{MN}$. Thus $R_1$ and
$R_2$ are
the radii of the circles, measured in the closed string metric, along
$x^1$ and $x^2$ directions
respectively.
With the normalization of $F_{\mu\nu}$ used in writing eq.\refb{e1},
the quantization law of the magnetic flux is given by:
\be \label{e3a}
F_{12}={n\over 2\pi}\, .
\ee
{}From eq.\refb{e1} we see that the mass of the wrapped
D2-brane is given by
\be \label{e4}
M_{D2} = {1\over g_c} \sqrt{R_1^2 R_2^2 + (2\pi F_{12})^2}
={1\over g_c} \sqrt{R_1^2 R_2^2 + n^2}\, .
\ee 

Now, according to the results of ref.\cite{9908142,9912274} we can
describe the
string
field theory living on this system around this background magnetic field 
by starting with a string field
theory written in a background metric $G_{\mu\nu}$, effective
coupling $g_o$, and zero background magnetic field, and then replacing all
products appearing in the action and the equations of motion of this
theory by
non-commutative $*$-products defined with 
non-commutativity parameter $\Theta^{\mu\nu}$. $G_{\mu\nu}$,
$\Theta^{\mu\nu}$ and $g_o$ are given in terms of $g_{\mu\nu}$,
$F_{\mu\nu}$ and $g_c$ as:
\be \label{e4a}
G^{-1} + {\Theta\over 2\pi} = (g+2\pi F)^{-1}\, , \qquad g_o =
g_c\sqrt{\det
G\over
\det(g+2\pi F)}\, .
\ee
This gives, using eqs.\refb{e3}, \refb{e3a}
\be \label{e5}
G = (R_1^2R_2^2 + n^2) \pmatrix{R_2^{-2} & \cr & R_1^{-2}}\, , \qquad
\Theta = {2\pi n\over R_1^2R_2^2 + n^2}\pmatrix{& -1\cr 1&}\, , \qquad g_o
=
g_c {\sqrt{R_1^2 R_2^2 + n^2}\over R_1 R_2
}\, .
\ee
{}From this we see that the radius of the $x^1$ direction measured in
the metric $G_{\mu\nu}$ is
given by $\sqrt{(R_1^2R_2^2 + n^2)/R_2^2}$. Let us adjust $R_1$, $R_2$ and
$n$
such
that this radius is unity, {\it i.e.}
\be \label{e6}
R_1^2R_2^2 + n^2 = R_2^2\, .
\ee
We shall now show that when eq.\refb{e6} is satisfied, there is a marginal
deformation which takes the CFT describing the D2-brane system under study
to a D1-brane along $x^2$. For this
let us consider an auxiliary system where the non-commutativity
parameter is set to zero, keeping $G_{\mu\nu}$ and $g_o$ fixed at values
given in eqs.\refb{e5}. This will
correspond to a D2-brane wrapped on $T^2$ with zero background magnetic
field, and closed string metric and coupling constant given by
$G_{\mu\nu}$ and $g_o$ defined in
eq.\refb{e5}. When eq.\refb{e6} is satisfied, the radius in the $x^1$
direction is unity for this auxiliary system, and the results of
refs.\cite{RECK,9902105} show that there is an exact marginal
deformation in the
boundary CFT describing this auxiliary system, generated by the operator
$\cos(X^1)$. Furthermore, for a
specific
value of the deformation parameter, the deformed CFT associated with this
auxiliary system represents a D1-brane lying along the
$x^2$ direction. In the language of string field theory, the existence of
this marginal deformation implies the
existence of a one parameter family of solutions, with fields depending on
the $x^1$ direction\cite{0007153}. 

Let us now go back to the original system, describing the D2-brane on
$T^2$ with $n$ units of magnetic flux on it. Equations of motion in the
string field theory describing this system differs from those in the
auxiliary system by the replacement
of all the ordinary products by $*$-products\cite{9908142,9912274}. But
the $*$-product reduces to the ordinary product for field
configurations which depend on only
one direction. Thus the one parameter family of solutions in the string
field theory describing the auxiliary system are also classical solutions
in the
string field theory describing the original system.

In the language of two dimensional conformal field theory, this means that
for the D2-brane on $T^2$ in the background given in eqs.\refb{e3},
\refb{e3a}, the boundary CFT admits
an exactly marginal perturbation when eq.\refb{e6} is satisfied, (Special
cases of such deformations for $R_1=R_2$ were discussed in
ref.\cite{9503014}.) 
One would naturally suspect that just as in the case of the auxiliary
system, this CFT also flows to a D1-brane along $x^2$ under this
marginal deformation.
That this is indeed so can be argued by noting that
the correlation functions of open string vertex
operators in the boundary CFT, which carry momentum only along the $X^1$
direction, are identical to that in the auxiliary CFT.
Thus
we can borrow the results of \cite{RECK,9902105} and
conclude
that there is a
special point along the direction of marginal deformation where the
Neumann boundary condition along $x^1$ gets converted to a
Dirichlet boundary condition. This gives a D1-brane along the $x^2$
direction.

There are various consistency checks that one can perform to confirm this
result: 
\begin{enumerate}
\item
The mass of a D1-brane wrapped along $x^2$ is given by
\be \label{e7}
M_{D1} = {R_2\over g_c}\, .
\ee
When eq.\refb{e6} is satisfied, the mass of
the D1-brane given in \refb{e7} matches with that of the D2-brane given in 
\refb{e4}.

\item
If
there is a marginal deformation interpolating between the initial D2-brane
configuration and the final D1-brane configuration, then the CFT
describing the final
D1-brane system must also admit a marginal deformation. In the analysis of
refs.\cite{RECK,9902105} this came from massless open string modes winding
around the
1 direction.
The ground state of such an open string has mass$^2$ equal to
\be \label{e7a}
(R_1^2-1)=
-n^2/R_2^2\, ,
\ee
using \refb{e6}.
This is strictly negative for $n\ne 0$, and describes a tachyonic mode
rather than a massless mode. This
poses a puzzle. However note that if we consider an open string state
winding once along the $x^1$ direction and carrying $n$ units of momentum
along the $x^2$ direction, then it has mass$^2$
\be \label{e8}
R_1^2 + {n^2 \over R_2^2} - 1 =0\, .
\ee
Thus in this case these open string modes give rise to the marginal
deformation which takes us back to the original D2-brane system with a
magnetic field on it. This calculation also illustrates the importance of
quantization law of $F$; if $n$ in eq.\refb{e6} had not been an integer,
the spectrum of
open strings on the D-string wrapped along $x^2$ would not contain a
massless state of this kind.

\item
The analysis  can be given a more intuitive interpretation in a T-dual
language. 
For this let us make an $R\to 1/R$ duality transformation along the $x^2$
direction so that the dual $\wt x^2$ direction now has radius $R_2^{-1}$.
In
that case the initial configuration describes a D-string pointing
along the vector $(1,n)$ in the $(x^1,\wt x^2)$ plane. The total length of
such a D-string inside a unit cell is $2\pi\sqrt{R_1^2 + n^2/R_2^2}$.
It can be easily
verified that when condition \refb{e8} is satisfied, the open string state
on this D-string, carrying unit momentum along the D-string, is exactly
marginal. 
Using the results of refs.\cite{9902105,RECK} we can show that this
marginal
deformation takes the D-string to a D0-brane. We can now go back to the
original description by 
a reverse $R\to 1/R$ duality transformation along the $\wt x^2$ direction.
This
gives us a D-string stretched along $x^2$.
\end{enumerate}

Thus we have established that there is a marginal deformation which
takes us from an initial configuration of D2-brane with $n$ units of
magnetic flux to a D1-brane lying along $x^2$. We shall now show that
there is a relevant deformation of the CFT which takes this D1-brane to a
D2-brane
wrapped on $T^2$, but with no magnetic flux. As seen from eq.\refb{e7a},
the ground state of an open string with
unit winding along the $x^1$ direction, and no momentum along the $x^2$
direction, represents a tachyonic mode, and hence a relevant deformation
of
the boundary CFT. We shall now investigate the effect of switching on this
relevant perturbation on the conformal field theory describing the
D-string. This is indeed a well studied problem, and can be recognised as
such by going to the T-dual description in which we make an $R\to (1/R)$
duality transformation along the $x^1$ direction. This takes the
D-string to a D2-brane wrapped on the dual torus, and the open string
states with unit winding along $x^1$ to open string states with unit
momentum along the dual $\wt x^1$ direction. As has been shown in
refs.\cite{9406125,0003101}, the effect of perturbation by open string
vertex operators carrying unit momentum along the $\wt x^1$ direction is
to take the D2-brane to a
D1-brane lying along $x^2$. Going back to the original description by
another $R\to 1/R$ duality transformation along $\wt x^1$, we see that the
final
configuration is a D2-brane wrapped on the original torus $T^2$,
{\it without any background magnetic field}.

Thus by a combination of marginal and relevant deformations we can take a
D2-brane wrapped on $T^2$ with {\it $n$ units of magnetic flux} to a
D2-brane
wrapped on $T^2$ with {\it no magnetic flux}. In open string field theory
the
effect of a marginal deformation can be represented by a continuous
deformation of string field configuration via {\it on-shell field
configurations}, whereas a relevant deformation of the kind discussed here
can be represented by a continuous deformation of string field
configuration via {\it off-shell field configurations}. (To see how
relevant deformation can be regarded as a continuous deformation via
off-shell field configuration, we can take the solution given in
ref.\cite{0005036}, and continuously deform each component of the string
field from 0 to
the final
value appropriate for the solution.) Thus the result of this paper implies
that in string field theory describing dynamics of a D2-brane wrapped 
on $T^2$, there is a continuous deformation via off-shell string field
configurations which can interpolate between a configuration with $n$
units of magnetic flux through $T^2$ and a configuration with 0 unit of
magnetic flux through $T^2$. This establishes the desired result.

One can give a slight twist to this tale by taking $R_1$ to be 1 instead
of the value given in eq.\refb{e6}. In this case the $x^1$ radius of the
auxiliary system, given by $\sqrt{1 + n^2 R_2^{-2}}$, is larger than 1,
and as a result the open string vertex operator with unit momentum along
the $x^1$ direction is relevant rather than marginal. Nevertheless, the
analysis of refs.\cite{9406125,0003101} tells us that under this relevant
deformation the auxiliary CFT flows to that describing a D-string along
$x^2$. The arguments given earlier then shows that the same must be true
also for the original system consisting of a D2-brane with $n$ units of
magnetic flux.

Now start with a different configuration, $-$ a D2-brane wrapped on the
same torus but without any magnetic flux. For $R_1=1$ the ground state of
the open string with unit momentum along $x^1$ is massless and describes
an exact marginal deformation which takes the D2-brane to a D1-brane lying
along $x^2$. Thus we see that starting with two different D2-brane
configurations, one with magnetic flux and one without magnetic flux, we
reach the same D1-brane configuration. This in turn shows that the
codimension one soliton formed by tachyon condensation on the original
D2-brane is independent of the magnetic flux on the brane. This is
precisely what is predicted according to the conjecture that we are trying
to verify.

We conclude with the following observations:
\begin{itemize}
\item The analysis of this paper can also be applied to the
D-brane anti-D-brane system or non-BPS D-brane in type II string
theories. Magnetic field on a non-BPS D-brane does not give rise to any
Ramond-Ramond (RR) charge, and so the possibility of changing this
magnetic field does not violate any conservation law. For a D-brane
anti-D-brane system only a special combination of the magnetic field on
the two branes can be switched on this way, $-$ the one which
does not give rise to any RR charge.\footnote{I would like to 
thank S. Minwalla for discussion on this
point.}
Thus there is again no conflict with conservation of RR
charges.

\item The main lesson learnt from the analysis of this paper is that
whereas the
description of the world-volume theory of the D-brane in terms of the
effective action involving tachyon and the massless fields captures many
of the important features, it fails to capture all the important
properties of the system. This point has already been advocated forcefully
in ref.\cite{0008033} in a different context. Here we see another
illustration of the same phenomenon. Field configurations which are
disconnected in the low energy effective field theory get connected to
each other in the full string theory. On the other hand, description of
the D-brane system in terms of conformal field theory is suitable for
studying condensation of tachyons into a lower dimensional soliton, but
not into the vacuum. Thus full-fledged string field theory seems to be the
only framework for studying all aspects of the problem. 
\end{itemize}

{\bf Acknowledgement}: I would like to particularly thank B. Zwiebach for
critical reading of the manuscript and many useful suggestions. I also 
wish to thank R. Gopakumar and S. Minwalla for discussions and comments on
the
manuscript, and W. Taylor for discussions.

\end{document}